\documentclass[10pt,journal,twocolumn]{IEEEtran}
\IEEEoverridecommandlockouts
\usepackage{caption}
\usepackage[utf8]{inputenc}
\usepackage{cite}
\usepackage{amsmath,amssymb,amsfonts,booktabs}
\usepackage{algorithmic}
\usepackage{graphicx}
\usepackage{textcomp}
\usepackage{xcolor}
\usepackage{enumerate}
\usepackage{epstopdf}
\usepackage{subcaption}
\usepackage{amsthm}

\usepackage{algorithm}
\usepackage{algorithmic}
\allowdisplaybreaks[4]

\def\BibTeX{{\rm B\kern-.05em{\sc i\kern-.025em b}\kern-.08em
	T\kern-.1667em\lower.7ex\hbox{E}\kern-.125emX}}
\begin{document}

\title{Fluid Antenna Systems Meet Low-Altitude Wireless Networks: Fundamentals, Opportunities,\\and Future Directions}

\author{
        Wenchao Liu,
        Xuhui Zhang,
        Chunjie Wang,
        Jinke Ren,
        Weijie Yuan,
        and Changsheng You

\thanks{
Wenchao Liu and Weijie Yuan are with the School of Automation and Intelligent Manufacturing, Southern University of Science and Technology, Shenzhen 518055, China (e-mail: wc.liu@foxmail.com; yuanwj@sustech.edu.cn).
}

\thanks{
Xuhui Zhang and Jinke Ren are with the Shenzhen Future Network of Intelligence Institute, the School of Science and Engineering, and the Guangdong Provincial Key Laboratory of Future Networks of Intelligence, The Chinese University of Hong Kong, Shenzhen, Guangdong 518172, China (e-mail: xu.hui.zhang@foxmail.com; jinkeren@cuhk.edu.cn).
}

\thanks{Chunjie Wang is with Shenzhen Institutes of Advanced Technology, Chinese Academy of Sciences, Shenzhen 518055, China, and also with University of Chinese Academy of Sciences, Beijing 100049, China (e-mail: cj.wang@siat.ac.cn)
}

\thanks{Changsheng You is with the Department of Electronic and Electrical Engineering, Southern University of Science and Technology, Shenzhen 518055, China. (e-mails: youcs@sustech.edu.cn).
}

\thanks{Corresponding author: Weijie Yuan}

}

\maketitle

\begin{abstract}
Low-altitude wireless networks (LAWNs) are widely regarded as a cornerstone of the emerging low-altitude economy, yet they face significant challenges, including rapidly varying channels, diverse functional requirements, and dynamic interference environments. Fluid antenna (FA) systems introduce spatial reconfigurability that complements and extends conventional beamforming, enabling flexible exploitation of spatial diversity and adaptive response to channel variations.
This paper proposes a novel architecture for FA-empowered LAWNs and presents a case study demonstrating substantial improvements in communication, sensing, and control performance compared to fixed-position antenna (FPA) systems. 
Key practical deployment considerations are examined, including mechanical design, position control, energy efficiency, and compliance with emerging industry standards. 
In addition, several future research directions are highlighted, including intelligent optimization, multi-function integration, and the exploration of novel low-altitude applications.
By integrating theoretical analysis with practical deployment perspectives, this paper establishes FA systems as a key enabler for adaptive, efficient, and resilient network infrastructures in next-generation LAWNs.
\end{abstract}

\section{Introduction}
In recent years, the low-altitude economy has experienced rapid growth, encompassing applications such as unmanned aerial vehicle (UAV)-enabled parcel delivery, emergency response, precision agriculture, and urban surveillance \cite{11131292}. This emerging paradigm is reshaping the interaction between aerial resources and ground infrastructure while accelerating the utilization of low-altitude airspace. Heterogeneous aerial platforms, including UAVs and other airborne systems, must operate safely, efficiently, and collaboratively in complex urban environments. These developments impose stringent demands on low-altitude wireless networks (LAWNs), which must support reliable high-capacity data transmission, real-time environmental sensing, and precise wireless control, concurrently \cite{yuan2025ground}. However, existing wireless infrastructure often fails to meet these requirements in the highly dynamic and interference-rich conditions of low-altitude environments.

Most existing LAWNs rely on fixed position antenna (FPA) arrays deployed at base stations (BSs) and access points. Although advanced beamforming techniques can enhance link quality, FPA-based LAWNs suffer from three critical limitations. 
First, their adaptability to dynamic environments is limited, as the mobility of UAVs frequently alters line-of-sight (LoS) and multipath conditions, while obstacles such as buildings and complex terrain can cause severe link degradation. 
Second, they face intense resource competition among communication, sensing, and control (CSC) functions, necessitating higher transmit power and increased number of antennas to meet performance demands. 
Third, they suffer from severe interference issues, as extending coverage using FPA arrays inevitably amplifies co-channel interference, especially in dense low-altitude airspace where a large number of aerial and ground users contend for limited spectrum resources, making interference management a critical challenge.

Fluid antenna (FA), also referred to as movable antenna (MA), offer a promising solution to these challenges by introducing an additional spatial degree of freedom (DoF) \cite{10753482,10906511}. Unlike conventional FPA systems, which adjust only the beamforming, FA systems can physically reposition antenna elements within a predefined mechanical range. This physical repositioning enables flexible beamforming, allowing for dynamic adjustment of beam direction, concentration of signal energy toward high-demand regions, and improved link quality and throughput \cite{10858129}. In addition, FA-empowered LAWNs enhance the feasibility of multi-function integration by exploiting spatial reconfigurability to support the joint optimization of CSC within a unified framework.

Motivated by the above, this paper presents a comprehensive study of FA systems as a key enabler for LAWNs. Specifically, we first conduct a systematic analysis of the challenges in LAWNs, which significantly limit performance and scalability, thereby motivating the exploration of new antenna architectures. 
Then, the fundamental principles and potential capabilities of FA systems are elaborated, highlighting their spatial flexibility and reconfigurability. 
Leveraging these advantages, a novel FA-empowered LAWN framework is proposed, in which FA arrays are deployed on BSs, UAVs, and other low-altitude platforms, integrating the network architecture, FA-enabled multi-function integration, multi-layer cooperation mechanisms, energy and resource management strategies, and new design issues. 
Finally, practical deployment considerations, future research directions, and representative application scenarios of FA-empowered LAWNs are also discussed.

\begin{figure*}[t]
	\centering
\fbox{\includegraphics[width=0.95\linewidth]{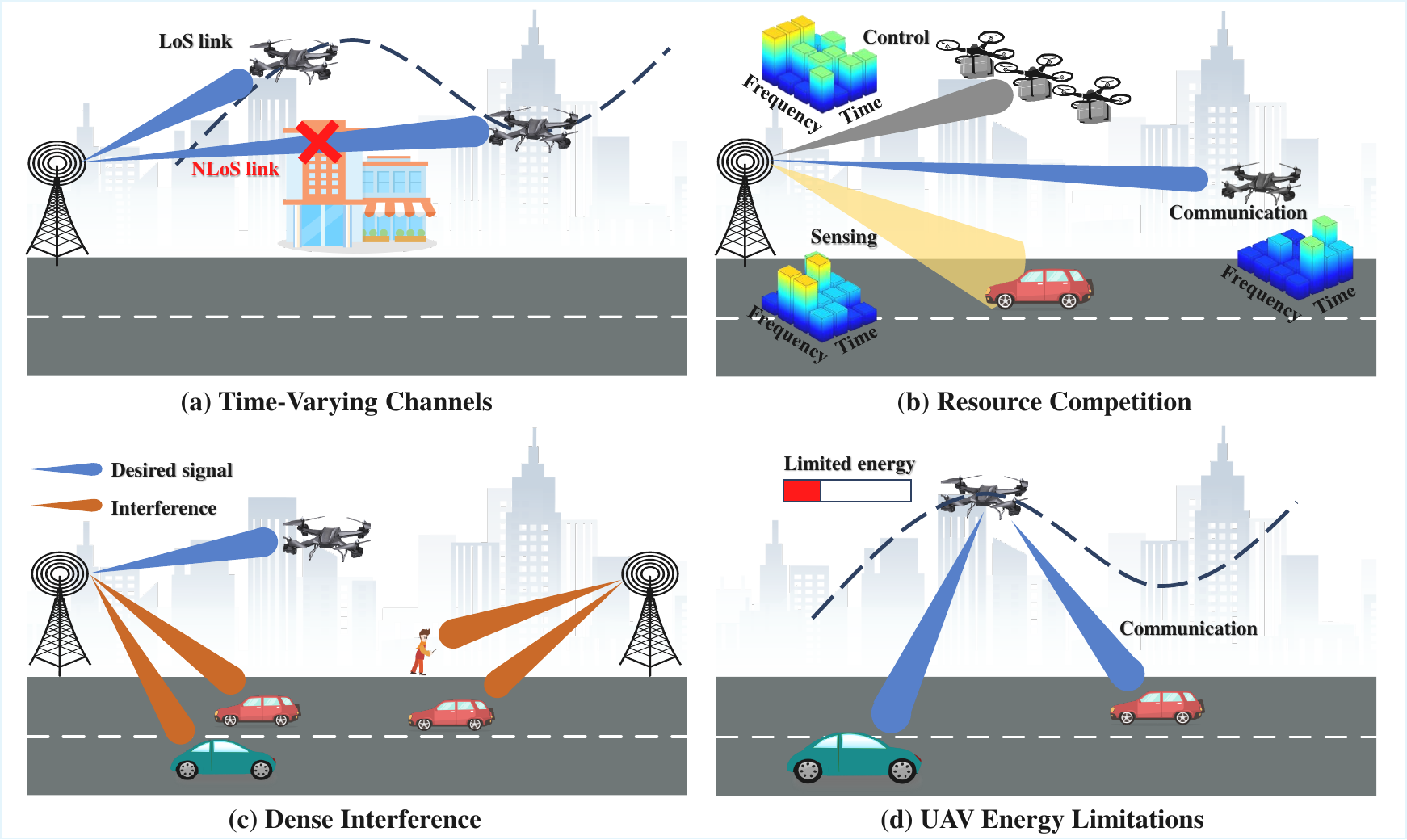}}
	\caption{Illustration of key challenges in LAWNs, including (a) time-varying channels, (b) resource competition, (c) dense interference, and (d) UAV energy limitations.}
	\label{fig:lawn challenge}
\end{figure*}

\section{Challenges in LAWNs}
LAWNs operate in a highly dynamic three-dimensional (3D) environment, where the high mobility of nodes and rapidly time-varying channels impose significant challenges on network design. Compared with conventional terrestrial networks, LAWNs differ fundamentally not only in mobility patterns but also in their requirement for highly integration of CSC, which imposes strict demands on reliability, scalability, and efficient resource utilization. Importantly, these challenges are not isolated but highly coupled: for instance, fast-varying channels make resource allocation more difficult, while LAWN performance is constrained by dense interference and limited energy resources. The challenges in LAWNs are described as follows.

\textbf{1) Time-Varying Channels:} 
Unlike conventional terrestrial networks, LAWNs are subject to rapidly changing channel conditions caused by multiple factors. The high mobility of nodes leads to frequent changes in position, altitude, and velocity, which in turn cause fast-varying Doppler shifts. LoS links are often blocked by urban structures, resulting in communication outages and severe fading, as illustrated in Fig. \ref{fig:lawn challenge}(a). Furthermore, reflections from buildings, ground surfaces, and other aerial objects lead to strong multipath effects. These factors make it difficult to obtain perfect channel state information (CSI), especially for latency-sensitive applications such as UAV flight control and real-time environmental sensing.

\textbf{2) Resource Competition:} 
LAWNs are required to simultaneously support three highly coupled functions: high-throughput communication for aerial-terrestrial data exchange, high-accuracy sensing for environmental monitoring and target detection, and high-reliability control for flight safety and UAV swarm cooperation. In practice, these functions inherently compete for limited wireless resources in the time, frequency, and spatial domains \cite{11045436}, as illustrated in Fig. \ref{fig:lawn challenge}(b). For instance, beamforming designs that improve sensing performance often reduce communication throughput, while spectrum allocated for reliable control signaling reduces the bandwidth available for data transmission. These trade-offs are further worsened by time-varying channels and interference, highlighting the need for joint multi-dimensional resource management schemes. 

\textbf{3) Dense Interference:} 
In the low-altitude environment, dense spectrum reuse across aerial and terrestrial systems causes severe interference. Interference sources include neighboring cellular cells, UAV swarms, and multi-domain communication links, which together create a complex 3D interference environment, as illustrated in Fig. \ref{fig:lawn challenge}(c). Conventional interference suppression techniques rely on the fixed spatial layout of FPA arrays, making them hard to adapt to rapidly changing interference conditions and dynamic network topologies. 

\textbf{4) UAV Energy Limitations:}
As UAVs form the core infrastructure of LAWNs, their limited battery capacity directly limits network endurance, making energy efficiency a critical issue, as illustrated in Fig.~\ref{fig:lawn challenge}(d). While propulsion usually accounts for the largest share of energy consumption, communication and sensing performance also significantly affects flight duration. In conventional FPA-based LAWNs, the lack of spatial reconfigurability often results in low data transmission rate, which results in long time duration to complete communication and sensing tasks. This longer operation time leads to higher propulsion energy use, thereby reducing task sustainability. 

\textbf{5) Limitations of Conventional Methods:} 
Several enhancement techniques have been explored for LAWNs, including massive multiple-input multiple-output (MIMO) beamforming, reconfigurable intelligent surfaces (RIS), and cooperative relaying schemes. However, massive MIMO brings high hardware complexity and limited adaptability to time-varying LoS blockages. RIS lacks active transmission capability and cannot effectively mitigate dynamic interference. cooperative relaying reduces link degradation but introduces high latency and energy overhead, especially in dynamic network conditions. These limitations highlight the need for more flexible and adaptive solutions designed for the unique characteristics of LAWNs. 

\section{Principles and Advantages of FA Systems}
By treating antenna position as an additional spatial DoF, FA systems extend beyond conventional beamforming and enable joint optimization of beamforming and antenna placement, leading to significant performance gains. In this section, we discuss the fundamental principles, typical architectures, core capabilities and advantages, and practical design considerations.

\textbf{1) Fundamental Principle:} 
In FPA systems, performance optimization is primarily achieved by adjusting beamforming, which determines the amplitude and phase applied to each antenna element, while the physical positions of the elements remain fixed. In contrast, FA systems can physically reposition antenna elements within a predefined movement region, thereby directly modifying the electromagnetic propagation geometry at the physical layer. From a signal processing perspective, the beamforming shapes the directional gain of the wavefront, whereas physical repositioning alters relative path lengths and channel correlations between the antenna elements and the users or targets. Through the joint optimization of beamforming and the FA positions, FA systems can synthesize radiation patterns that cannot be achieved by the FPA systems due to structural constraints, thereby enhancing spatial resolution, improving interference suppression, and increasing adaptability to dynamic environments.

\begin{figure}[t]
\centering
\fbox{\includegraphics[width=0.95\linewidth]{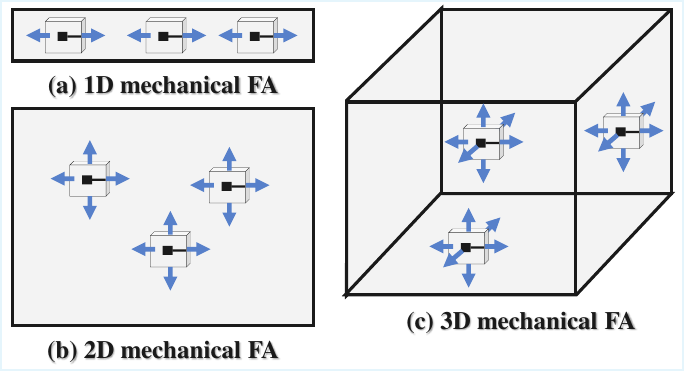}}
\caption{Illustration of mechanical FA architectures, including (a) 1D mechanical FA arrays, (b) 2D mechanical FA arrays, and (c) 3D mechanical FA arrays.}
\label{fig:FA}
\end{figure}

\textbf{2) Typical Architectures:}
The concept of FAs encompasses a broad class of reconfigurable and flexible antenna structures. Among the various implementations, one of the most representative approaches is the mechanical FA, which is also called MA \cite{10906511}, where a stepper motor driven mechanical slider is employed to reposition the antenna elements dynamically \cite{10753482}. Such mechanical FA architectures can generally be classified into three cases, as illustrated in Fig. \ref{fig:FA}.
One-dimensional (1D) mechanical FA arrays allow antenna elements to move along a straight path, either horizontally or vertically, enabling simple mechanical control and making them well suited for compact BS deployments and for integration into UAV platforms. 
Two-dimensional (2D) mechanical FA arrays allow antenna elements to move within a 2D plane, offering greater spatial flexibility and supporting adaptive coverage shaping in areas with high user density or changing traffic demands.
3D mechanical FA arrays provide full control over spatial position, which offer maximum spatial DoF and are particularly beneficial in complex environments such as dense urban areas or highly dynamic scenarios. For all these cases, FA position design must be closely integrated with motion control algorithms to effectively balance antenna movement speed, position accuracy, and energy efficiency.

\textbf{3) Core Advantages:} 
The physical repositioning of antenna position provides additional DoFs, which can effectively support the core performance requirements of LAWNs in highly dynamic and interference-rich environments. First, flexible beamforming is achieved by jointly adjusting antenna positions and beamforming, enabling transmission energy to dynamically focus on regions of high demand \cite{zhang2025movable}. Second, interference mitigation is realized through antenna repositioning, which modifies the interference geometry, steers nulls toward dominant interferers, and reduces spatial correlation, thereby significantly improving interference suppression \cite{10948162}. Third, antenna repositioning enhances the separability of user channel vectors in low-scattering environments, thereby improving spatial multiplexing gain and significantly increasing system throughput and spectral efficiency without requiring additional spectrum resources \cite{10753482}. Finally, through dynamic antenna repositioning, FA systems provide the capability to balance performance among CSC tasks, supporting efficient joint optimization under the highly dynamic conditions of LAWNs \cite{10130117}.

\textbf{4) Practical Design Considerations:} 
Although FA systems outperform conventional FPA systems by exploiting spatial reconfigurability, their performance relies strongly on precise antenna motion control and efficient real-time schedule strategies. The joint design of the antenna positions and the beamforming typically leads to non-convex optimization problems, which often require heuristic methods such as particle swarm optimization (PSO) algorithm \cite{10693833}. Furthermore, physical repositioning introduces mechanical latency that is slower than electrical beam switching, making scheduling strategies essential to minimize the overhead of frequent movements \cite{yu2025predictive}. Energy consumption is also a critical concern, especially in the UAV-enabled systems where actuation draws from limited onboard supplies. These practical challenges significantly influence the design and deployment of FA systems in LAWNs and will be further analyzed in Section \ref{Practical Deployment}.

\begin{figure*}[t]
	\centering
\fbox{\includegraphics[width=0.95\linewidth]{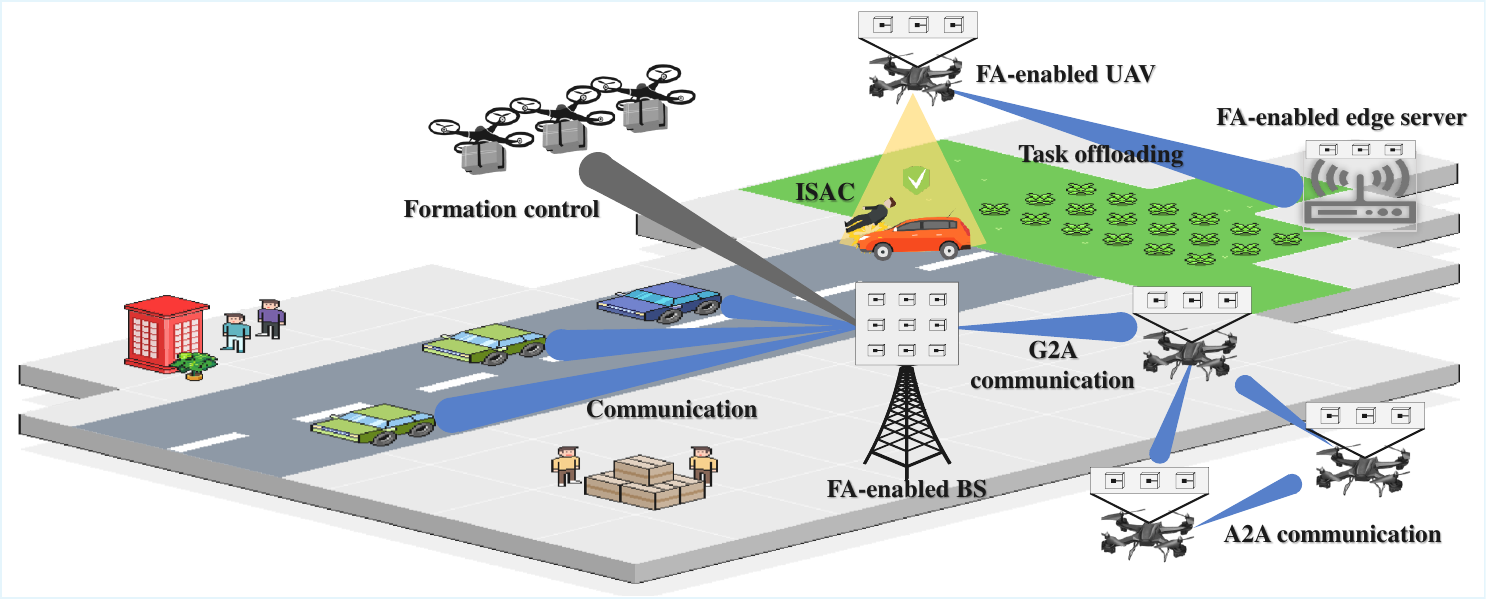}}
	\caption{Illustration of the system architecture for the FA-empowered LAWN, where the FA-enabled UAVs and the BSs collaboratively support multiple functions including integrated sensing and communication (ISAC), the A2A communication, the G2A communication, and formation control.}
	\label{fig:ma enabled lawn}
\end{figure*}

\section{FA-Empowered LAWNs}
Integrating FA systems into LAWNs requires a comprehensive architectural design to meet the operational requirements of aerial platforms, support diverse task scenarios, and adapt to the highly dynamic low-altitude environment. This section presents a novel architecture for FA-empowered LAWNs, emphasizing the integration of CSC functions, and describes the key system components needed to ensure network adaptability and reliability.

\textbf{1) Architectural Design:} 
An FA-empowered LAWN can be modeled as a layered, functionally integrated system architecture consisting of three core subsystems, as illustrated in Fig. \ref{fig:ma enabled lawn}.
The ground layer includes ground BSs and edge nodes integrated with FA arrays, which provide CSC services to both aerial and ground users.
The aerial layer consists of UAVs or other aerial platforms equipped with FA arrays to support air-to-air (A2A) and air-to-ground (A2G) communication links, while also carry sensor payloads for environmental sensing.
The control and management layer is centered on a software-defined networking (SDN) controller and network orchestrator, enabling centralized control and intelligent scheduling of global resource allocation, FA position movement, beamforming design, and joint scheduling of CSC tasks.

\textbf{2) Functional Integration of CSC:} 
FA systems enable LAWNs to dynamically adapt resource allocation based on task requirements, achieving adaptive multi-functional services over a shared physical infrastructure. 
For communication scenarios, A2A and A2G links are enhanced through adaptive beamforming, where FA arrays are repositioned to maintain favorable channel conditions \cite{10654366}. FA-enabled UAV relays establish cooperative A2A connections by forming dynamically adjustable narrow beams, improving inter-UAV connectivity while reducing interference with ground users.
In sensing scenarios, FA-empowered LAWNs enhance sensing capabilities by enabling precise beam steering and dynamic antenna repositioning \cite{11155198}. The ability to adjust antenna positions improves angular resolution, expands scanning coverage, and increases spatial diversity, thereby significantly enhancing sensing performance in challenging low-visibility conditions such as fog or smoke.
For control scenarios, low-latency and highly reliable links are essential, particularly for UAV formations operating in complex urban environments. FA-empowered LAWNs can achieve significantly reliable control performance by dynamically adjusting antenna positions, ensuring robust connectivity even under highly dynamic conditions. Building on this capability, antenna repositioning can be exploited to prioritize critical control or emergency signals, thereby enhancing the resilience and safety of system operations during faults or hazardous flight situations.

\textbf{3) Multi-Layer Cooperation:} 
The full potential of FA systems in LAWNs can be realized through multi-layer cooperation. At the physical layer, antenna positions can be jointly optimized with beamforming to adapt to instantaneous channels. At the medium access control (MAC) layer, CSC tasks can be dynamically scheduled according to channel quality, task priority, and urgency, with performance dependent on the reliability of channel prediction and resource management. At the network layer, SDN-based orchestration can coordinate aerial and ground nodes to achieve global resource optimization, contingent upon efficient CSI collection and low-latency feedback.

\textbf{4) Energy and Resource Management:} 
Although FA systems can significantly improve link quality and spectral efficiency, their physical repositioning introduces additional mechanical energy consumption, which becomes a critical issue for energy-constrained subsystems. To address this, the proposed architecture incorporates an energy-aware FA position scheduling mechanism that triggers repositioning only when the expected performance gain exceeds the associated energy cost.

\textbf{5) New Design Issue:} 
Compared with conventional LAWNs that rely on FPA systems, the introduction of FA systems brings entirely new DoFs as well as design challenges. In static scenarios, such as communication systems involving stationary UAVs or fixed BSs, it is necessary to jointly optimize the position of FAs and their transmit/receive beamforming. Although large-scale repositioning of FAs can actively search for more favorable channel conditions, the physical movement of the FAs introduces a non-negligible response delay. This movement delay directly reduces the available time duration for communication services. Particularly in latency-sensitive scenarios (e.g., real-time control and emergency communications), the movement delay can significantly reduce system throughput. Therefore, a trade-off must be considered between the channel gain improvements brought by antenna mobility and the communication performance loss caused by movement delay.

In more complex dynamic scenarios, such as highly dynamic UAV-based networks, the system must jointly optimize the UAV trajectory, the FA positions, and the transmit/receive beamforming. In this scenarios, the movement frequency of antenna positions becomes a critical design parameter: the FA positions frequent movement help continuously track fast time-varying channels, enhancing beamforming gain and link stability, but at the cost of higher energy consumption. Moreover, the movement of the antennas may introduce additional Doppler shifts and phase disturbances, thereby increasing the signal processing burden at the receiver. Consequently, the system faces multi-objective trade-offs among the movement frequency of FA positions, communication performance, and energy efficiency .

\textbf{Summary:} 
The proposed FA-empowered LAWN architecture achieves seamless integration of CSC by exploiting the spatial reconfigurability of FA systems in both aerial and ground nodes. This design effectively addresses environmental dynamics, interference management, and resource competition in LAWNs. 

\section{Practical Deployment Considerations} \label{Practical Deployment}
While simulations demonstrate the performance benefits of FA-empowered LAWNs, realizing these benefits in real-world requires careful consideration of multiple engineering and operational challenges. These challenges can be grouped into four key areas: mechanical design, position control, energy efficiency, and standard compliance.

\textbf{1) Mechanical Design:} 
The physical design of mechanical FA systems is critical to their long-term reliability. Unlike FPA systems, FA systems utilize stepper motor driven mechanical sliders to achieve dynamic repositioning of antenna elements. However, repeated mechanical actuation over time may lead to rail wear, joint loosening, and positional drift. To ensure durability, practical mechanical FA designs should incorporate high-precision actuators, low-friction drive components, and sealed enclosures to mitigate performance degradation caused by moisture, dust, and other environmental factors.

\textbf{2) Position Control:} 
The effectiveness of mechanical FA systems depends on the accuracy with which their physical position aligns with the intended target. At frequencies below 6 GHz, centimeter-level precision is often sufficient. However, millimeter-wave and terahertz bands require sub-millimeter accuracy to avoid beam misalignment and distortion. Closed-loop feedback using encoders or optical sensors can compensate for mechanical errors and mitigate the impact of environmental disturbances. In highly dynamic scenarios, predictive control is essential. By forecasting the motion states of UAVs or users, the antenna positions can be proactively adjusted to maintain stable performance despite actuation delays.

\textbf{3) Energy Efficiency:} 
Although mechanical FA systems can reduce the transmit power required to maintain high link quality, dynamic adjustment of antenna positions leads to additional energy consumption due to mechanical movement. In energy-constrained subsystems, this overhead constitutes a critical design challenge. To address this, energy-aware scheduling mechanisms should be employed that trigger reconfiguration only when the expected performance gain outweighs the energy cost of movement and incorporate motion aggregation to minimize movement frequency. For ground BSs, where energy constraints are less stringent, scheduling should still be optimized in dense deployment scenarios to reduce cumulative energy usage and lower operational costs.

\textbf{4) Standard Compliance:} 
Finally, LAWNs integrated with FA systems must comply with industry standards. Compatibility with 3GPP Release 18 and later, along with support for ongoing ISAC standardization efforts, will facilitate smooth network evolution and seamless integration \cite{he2025ubiquitous}. Standard compliance not only ensures interoperability with existing infrastructure but also guides the development of FA systems under unified specifications, thereby reducing commercialization risks. Furthermore, a hybrid deployment strategy that introduces FA systems selectively at critical network nodes can significantly reduce initial investment and enable a gradual transition toward FA-empowered LAWNs.

\section{Case Study: FA-empowered LAWNs}
This section presents a representative case study that encompasses the system model, the proposed solution, and the corresponding numerical results. The analysis evaluates the performance benefits of the proposed FA systems over conventional FPA systems in the CSC tasks.

\subsection{System Model and Proposed Solution}
To ensure generality, the case study considers independent communication users, sensing targets, and controlled UAVs. The BS simultaneously transmits communication signals to $K$ users, sensing signals to $M$ targets, and control commands to $N$ controlled UAVs, and is equipped with an FA array that can move within a 2D region. 

The objective is to minimize the BS transmit power by jointly optimizing the transmit beamforming and the FA positions, subject to the communication rate requirements for users, the sensing beampattern requirements for targets, and an acceptable linear quadratic regulator (LQR) cost \cite{11159297} for UAV control \footnote{In control theory, the LQR cost is used to quantify the “cost” or “loss” associated with system control performance. Its core role is to capture the overall trade-off between the deviation of the system state from the desired value and the energy consumed by the control input.}. 
This non-convex problem is challenging due to the strong coupling between the beamforming and the FA positions. To tackle it, a two-layer PSO-based algorithm is introduced, i.e., the inner layer applies successive convex approximation (SCA) to optimize beamforming with fixed FA positions, while the outer layer updates FA positions using the PSO algorithm, enabling adaptive spatial reconfiguration that simultaneously enhances CSC performance.

\begin{figure*}[t]
	\centering
	\fbox{\includegraphics[width=0.95\linewidth]{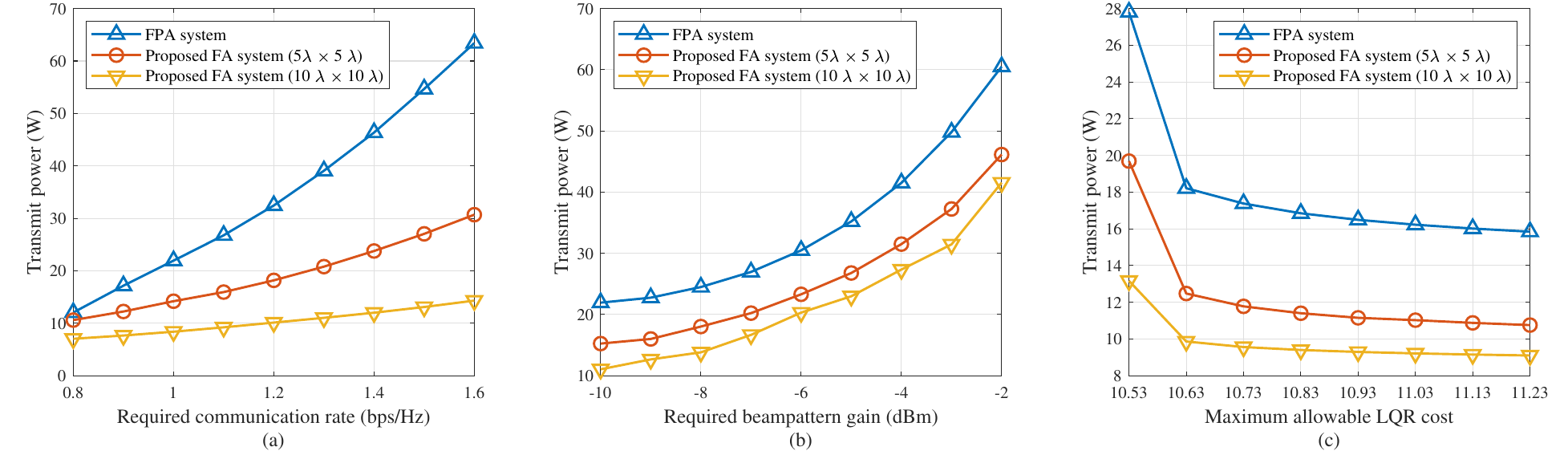}}
	\caption{Simulation results of the case study: (a) The transmit power versus the required communication rate, (b) The transmit power versus the required beampattern gain, and (c) The transmit power versus the maximum allowable LQR cost.}
	\label{fig:simulation4}
\end{figure*}

\subsection{Numerical Results and Discussion}
For performance comparision, a conventional FPA system is considered as the baseline.
Unless otherwise stated, numerical parameters are set as follows: the number of communication users is $ K = 3 $, the number of sensing targets is $ M = 3 $, and the number of controlled UAVs is $ N = 2 $. The BS is equipped with $ T_x = 10 $ transmit antennas. The noise power at the receiver is -100 dBm, and the channel power gain at the reference distance of 1 m is -60 dB. The required communication rate per user is 1 bps/Hz, the required beampattern gain toward the target direction is -10 dBm, and the maximum allowable LQR cost for the controlled UAVs is 10.58.

Fig. \ref{fig:simulation4} illustrates the transmit power performance under different system requirements. 
As shown in Fig. \ref{fig:simulation4}(a), the transmit power increases with the required communication rate for all considered systems. The FPA system shows a sharp increase, indicating that a significant increase in transmit power is required to meet the high-rate communication demands without the capability to reposition antennas. In contrast, FA system reduces transmit power significantly by repositioning antennas within the movable region to enhance channel gain. Moreover, enlarging the movable region (from $5\lambda \times 5\lambda$ to $10\lambda \times 10\lambda$) provides greater spatial DoFs, resulting in lower transmit power requirements.
Fig. \ref{fig:simulation4}(b) shows that the transmit power grows rapidly as the required beampattern gain increases. Realizing stronger beampattern gain typically requires more concentrated beams or more favorable phase/position, resulting in nonlinear power growth. FA systems achieve the same beamforming gain with lower transmit power by flexibly adjusting antenna positions, while larger movable regions further improve beam synthesis and yield higher power savings, particularly in high-gain regimes. 
Fig. \ref{fig:simulation4}(c) depicts the transmit power with respect to the maximum allowable LQR cost. When the control demand is stringent (i.e., the allowable LQR cost is small), all systems exhibit relatively high transmit power. As the control demand is gradually relaxed (i.e., the allowable LQR cost increases), the required transmit power of all systems decreases rapidly and eventually stabilizes. It is noteworthy that the FA system with a larger movable region are less sensitive to control requirements, i.e., under the same allowable LQR cost, an FA system with a larger movable region can satisfy the control demand using lower transmit power.

\section{Future Research Directions and Applications}
FA systems in LAWNs remain in the early stages of research. While the potential benefits of FA systems have been demonstrated, several emerging research directions in LAWNs are expected to further broaden FA's application scope. These opportunities can be categorized into three major areas: intelligent control and optimization, multi-function and multi-domain integration, and novel low-altitude application scenarios.

\textbf{1) Intelligent Control and Optimization:} 
To fully exploit the spatial reconfigurability of FA systems, future LAWNs will require intelligent and predictive control mechanisms capable of adapting to dynamic environments in real time. Conventional optimization approaches, which alternately optimize beamforming and antenna position, often suffer from excessive latency or high computational complexity. A promising solution is the integration of long short-term memory (LSTM)-based channel prediction with deep reinforcement learning (DRL) to enable adaptive FA position and beamforming strategies, allowing optimal antenna positions to be obtained in advance. Moreover, when integrated with digital twin platforms that provide virtual replicas of the network, such approaches facilitate large-scale pre-deployment testing and over-the-air optimization updates. 

\textbf{2) Multi-Function and Multi-Domain Integration:} 
LAWNs are expected to increasingly integrate with satellite and maritime systems, opening up numerous opportunities for cross-domain applications and enabling more flexible, resilient, and high-capacity connectivity. In this evolving landscape, FA systems emerge as a key enabler, offering distinct advantages over conventional FPA systems. For example, by dynamically adjusting antenna positions, FA-enabled integrated sensing and communication can simultaneously enhance data rates and improve sensing accuracy, supporting more reliable and efficient network operations. Furthermore, FA-assisted physical layer security provides an additional layer of protection by adaptively adjusting antenna positions to disrupt potential eavesdroppers, thereby improving the secrecy communication rate.

\textbf{3) Emerging Low-Altitude Application Scenarios:} 
As the low-altitude economy grows, FA-empowered LAWNs are expected to play a key role. In urban air mobility, FA systems can provide reliable, low-latency communication and accurate sensing for passenger UAVs and air taxis navigating complex 3D city environments. In disaster response, UAV swarms equipped with FA arrays can quickly restore connectivity, perform detailed environmental mapping, and cooperate autonomous rescue operations in areas where ground infrastructure is damaged or unavailable. In precision agriculture and environmental monitoring, UAV equipped with FA arrays can deliver both high-speed communication links and fine-grained sensing capabilities, supporting tasks such as crop health monitoring, pollution tracking, and wildlife observation. Across these applications, the main advantage of FA systems is their ability to adapt dynamically to specific task requirements, optimizing coverage, reducing interference, and improving resource efficiency while maintaining a flexible and unified network architecture.

\section{Conclusions}
LAWNs face significant challenges arising from rapidly varying channel conditions, stringent multi-functional requirements, and highly dynamic interference environments. This paper presents FA systems as a promising solution, highlighting its spatial reconfigurability as a complement to conventional beamforming to enhance CSC.
Through a case study, it is demonstrated that FA-enhanced LAWNs achieve significant improvements in CSC performance compared to conventional FPA systems. Beyond simulation analysis, this paper also examines key practical deployment considerations, including mechanical design, position control, energy efficiency, and compliance with emerging industry standards.
By integrating theoretical analysis with practical insights, this paper lays the foundation for developing adaptive and intelligent LAWNs. FA systems represents a critical advancement toward the realization of robust, efficient, and secure networks capable of meeting the growing demands of future low-altitude services.

\bibliographystyle{IEEEtran}
\bibliography{myref}
\end{document}